\documentclass[apj]{emulateapj}
\begin{document}
\bibliographystyle{apj}

\title{First order particle acceleration in magnetically-driven flows}

\author{Andrey Beresnyak}
\affil{Naval Research Laboratory, Washington, DC 20375}
\author{Hui Li}
\affil{Los Alamos National Laboratory, Los Alamos, NM, 87545}

\begin{abstract}
We demonstrate that particles are regularly accelerated while experiencing curvature drift in flows driven by magnetic tension.
Some examples of such flows include spontaneous turbulent reconnection and decaying magnetohydrodynamic (MHD) turbulence, where magnetic field relaxes
to a lower-energy configuration and transfers part of its energy to kinetic motions of the fluid.
We show that this energy transfer which normally causes turbulent cascade and heating of the fluid,
also results in a first-order acceleartion of non-thermal particles.
Being very generic, this acceleration mechanism is likely to be responsible
in production of non-thermal particle distribution in magnetically-dominant environments such as solar chromosphere, pulsar magnetosphere,
jets from supermassive black holes and $\gamma$-ray bursts.
\end{abstract}

\maketitle

\section{Introduction} 
Magnetically-dominated environments are fairly common in Astrophysics and a sizable fraction of astrophysical objects, that we know, is made of extremely rarefied plasma. These objects are only visible to us because they contain a non-thermal particle population which dominates emission across most of the objects spectrum.
A basic idea that particles must be accelerated, therefore, has been around for some time. One of the main
elements of the explanation of how energy gets to particles involves generic mechanisms of how energy
may be dissipated in an almost inviscid and perfectly conducting medium, i.e. how energy is transferred to small
scales where is is available to particles. Of these mechanisms the most popular are three: a) discontinuities
in the fluid motion e.g. shocks, b) turbulence and c) dicontinuities in the magnetic field, e.g. current sheets.
Following the idea of \citet{fermi1949} that collisionless particles can get energy by scattering in fluid motions, especially converging motions, the diffusive shock acceleration mechanism (DSA) \citep{krymskii1977,bell1978,malkov2001} has become rather popular in explaining non-thermal emission,
as well as energetic charged paricles, detected at Earth, known as cosmic rays. 
The acceleration rate, however, is related to the scattering rate, which is critically bound by the Bohm limit,
which makes acceleration to higher energies slower and slower.

The observations of some variable astrophysical objects, however, suggested extremely fast acceleration timescales, incompatible with diffusive shock acceleration.
Some radio jets powered by supermassive black holes exhibit $\sim 10$ min variations in TeV emissions, e.g., \citet{aharonian2007,aleksic2011}. Such fast time variabilities, along with other emission region constraints, have been argued in \cite{Giannios2013} as being evidence for mini-jets generated by reconnection.
Recent observations of gamma-ray flares from Crab \citep{abdo2011}
have suggested that the impulsive nature of the energy release and the associated particle acceleration might need an alternative explanation as well \citep{Clausen2012}. It has also been suggested that reconnection plays a crucial role in producing high energy emissions from gamma-ray bursts \citep{Zhang2011}.
As far as magnetically dominated enviroments are concerned, e.g., the solar corona or the pulsar wind nebula, it is
natural to expect the major source of energy to be magnetic energy, which is why reconnection and associated phenomena has been a active field of study, see, e.g. \citet{Uzdensky2015} for a review.
It would be especially interesting to find if there a generic mechanism to transfer magnetic energy into particles
and to conceptually understand the nature of recent numerical results that demonstrated that in both
MHD fluid simulations \citep{kowal2012b} and ab-initio plasma simulations \citep{Sironi2014,Guo2014,Uzdensky2014} of reconnection
there is a regular acceleration of particles

Particle acceleration is often classified into ``first order Fermi'' mechanism where particles are gaining energy
regularly, e.g., by colliding with converging magnetic mirrors and ``second order Fermi'' where particles can
both gain and lose energy \citep{fermi1949}. These two are not mutually exclusive and represent two different terms in the equation for the evolution of the distribution function: the term describing advection in energy space
and the term describing diffusion in energy space. Speaking of practical applications, the first order mechanism usually dominates, if present. The outcome of acceleration
depends on the rate at which particle gain energy, called acceleration rate $r_{acc}$ and the rate of particle escape from the system,
$r_{esc}$. If escape is negligible and $r_{acc}$ is constant with energy, the energy grows exponentially. Also,
if $r_{esc}/r_{acc}$ does not depend on energy, the stationary solution for the particle distribution is a power-law, with the power law index
determined by $-1-r_{esc}/r_{acc}$, see, e.g. \citet{drury1999}.
Various environments, such as supernova shocks, were thought to satisfy this condition and produce power-law distributed cosmic rays, which become
consistent with observations after being modified by diffusion. Acceleration within many orders of magnitude in energy
was regarded as a result of a large-scale physical layout of the acceleration site, e.g., the planar shock can be thought
of as a set of large-scale converging mirrors. The very same picture could also be applied to the large-scale reconnection site, where the two sides of the inflow
effectively work as converging mirrors \citep{dalpino2005}. In this paper we deviate from this mindset 
of the problem that achive scale-free acceleration just because there is only a single scale --
the scale of the system. Instead we will try to find regular acceleration
over large energy ranges in systems that do not necessarily possess global regular structure -- however
they could still be scale-free in statistical sense, such as turbulent systems. Normally, turbulent environments are expected to be regions of second-order acceleration, see, e.g., \citet{Schlickeiser2002,Cho2006}.
In this paper we point to the mechanism of regular or first order acceleration that was overlooked in the literature. This mechanism is inherently related to a certain statistical measure of energy transfer
in turbulence and, therefore, does not rely on a particular geometry and is very robust. As we will
show below, the direction of energy transfer from magnetic field to kinetic motions and the sign of curvature drift acceleration are inherently
related, so that in systems with the average positive energy transfer from magnetic energy to kinetic motions there is an average positive
curvature drift acceleration, while in the opposite case, there is an average curvature drift cooling.

\begin{figure}
\includegraphics[width=1.0\columnwidth]{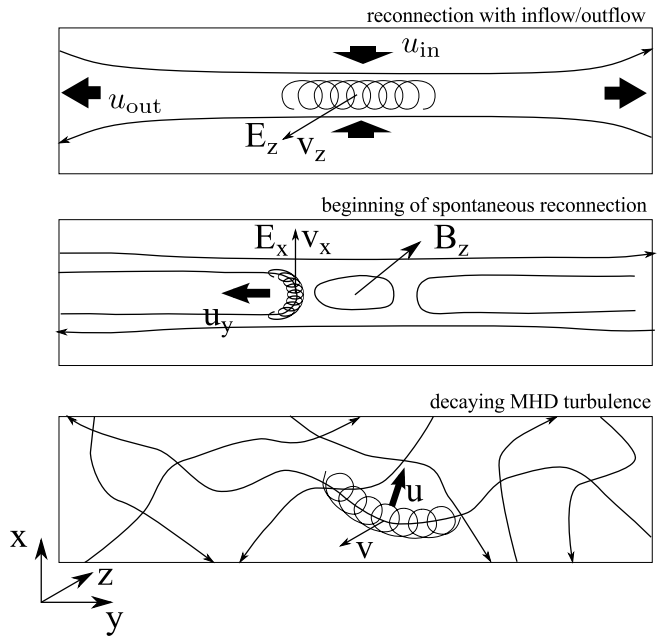}
\caption{A cartoon of several acceleration mechanisms in magnetized environments.
 The upper panel depicts reconnection with inflow and outflow where particles can be accelerated regularly due to the gradient drift and the large-scale electric field 
 The acceleration term is dominated by $v_z$ and $E_z$.
 The middle panel depicts initial stages of spontaneous reconnection which has negligible inflow,
 therefore the gradient drift term averages out. 
 In this case acceleration is mostly driven by contracting field lines which drive fluid motion and at the same time cause the curvature drift of particles.
 Note that the direction of drift is typically perpendicular to the fluid motion. The dominant acceleration term is associated with $E_x$ and the field curvature of $B_y$ component.
 The bottom panel depicts the same mechanism in a more homogeneous and isotropic setup of decaying MHD turbulence, which also has contracting field lines.
 In this situation all vector components contribute equally.}
\label{reconn}
\end{figure}

One of the commonly considered cases of magnetically-driven flows is magnetic reconnection. 
A significant effort was put into understanding on non-ideal plasma effects that could both cause reconnection and create non-zero parallel component
of the electric field \citep{pritchett2006,fu2006}. While non-ideal effects are indeed required for the individual field lines to break and reconnect,
their influence is limited to fairly small scales, typically below the ion skin depth $d_i$. In this paper, we instead focus on larger scales and ignore non-ideal effects for the following two reasons. Firstly, it has been argued that understanding of global energetics of large-scale reconnection, such as the amount of magnetic energy dissipated
per unit time, does not necessarily require detailed knowledge on how
individual field lines break and reconnect \citep{Lazarian1999,Eyink2011b,loureiro2007,B13a}. These global energetic
parameters could be more important for acceleration to high energies than local non-ideal effects. 
Secondly, in order to understand high energy particle acceleration, one normally has to consider plasma dynamics on scales much larger than $d_i$, that is on MHD scales. While modern simulations such as \citet{Sironi2014,Guo2014,Uzdensky2014},
are able to reach box sizes of several hundred $d_i$, some theory work is needed to disentangle contribution from MHD field and the non-ideal field to acceleration.

\section{MHD flows and energy transfer} 
Well-conductive plasmas can be described on large scales as inviscid and perfectly conducting fluid (ideal MHD). The ideal MHD equations allows for exchange between
thermal, kinetic and magnetic energies.
The Lorentz force density, multiplied by the fluid velocity, ${\bf u \cdot [j\times B]/}c$ is the amount of energy transferred from 
magnetic to kinetic energy.
While macroscopic (i.e. kinetic plus magnetic) energy is expected to be conserved in the ideal MHD, it is not the case
in real systems which have non-zero dissipation coefficients. This is qualitatively explained by the nonlinear turbulent cascade
that brings macroscopic energy to smaller and smaller scales until it dissipates into thermal energy. One of the important examples of this is the spontaneous reconnection where 
the thin current layer becomes turbulent and starts dissipating magnetic energy at a constant rate.
The small scales of these turbulent flows resemble ``normal'' MHD turbulence which has equipartition between
magnetic and kinetic energies \citep{B13a}, it is also true that kinetic and magnetic part of the cascade each contribute around half of the total cascade rate. Therefore, if we assume that the turbulent cascade is being fed with magnetic energy, approximately half of the magnetic energy has to be transferred into kinetic energy before equipartition cascade sets in. It follows that the 
$\bf B$ to $\bf v$ energy transfer must be {\it positive on average} and could be approximated by
one half of the volumetric energy dissipation rate $\epsilon$, the main parameter of turbulence.
The term ${\bf u \cdot [j\times B]/}c$ is the Eulerian expression for the work done by magnetic
tension upon the fluid element. This term can be rewritten as the sum of $-{\bf (u\cdot\nabla)} B^2/8\pi$, advection of magnetic energy density by the fluid, and 
\begin{equation}
{\cal T}_{bv}={\bf u \cdot (B\cdot \nabla)B}/4\pi,
\end{equation}
the actual energy transfer between $\bf B$ and $\bf v$.
For the purpose of future calculations we will separate the ${\cal T}_{bv}$ in the following way:

\begin{eqnarray}
{\cal T}= \frac{1}{4\pi}{\bf u \cdot (B\cdot \nabla)B}=\nonumber \\
 \frac{1}{4\pi} {\bf (u \cdot B)(b \cdot \nabla)}B+\frac{B}{4\pi} {\bf u \cdot (B\cdot \nabla)b} = \cal X +\cal D,
\end{eqnarray}

where we designated ${\bf b=B}/B$, a unit magnetic vector.
The term $\cal X$ contains cross helicity density ${\bf u \cdot B}$.
We argue that in those systems where $\langle{\bf u \cdot B}\rangle=0$, which include many physically relevant
cases, such as spontaneous reconnection, the whole $\cal X$ term could average out (see also Fig.~2).
The second term $\cal D$ contains magnetic field curvature ${\bf (B\cdot \nabla)b}$ and will be important for subsequent calculation of curvature drift. 


\begin{figure*} [h!t] 
\includegraphics[width=\textwidth]{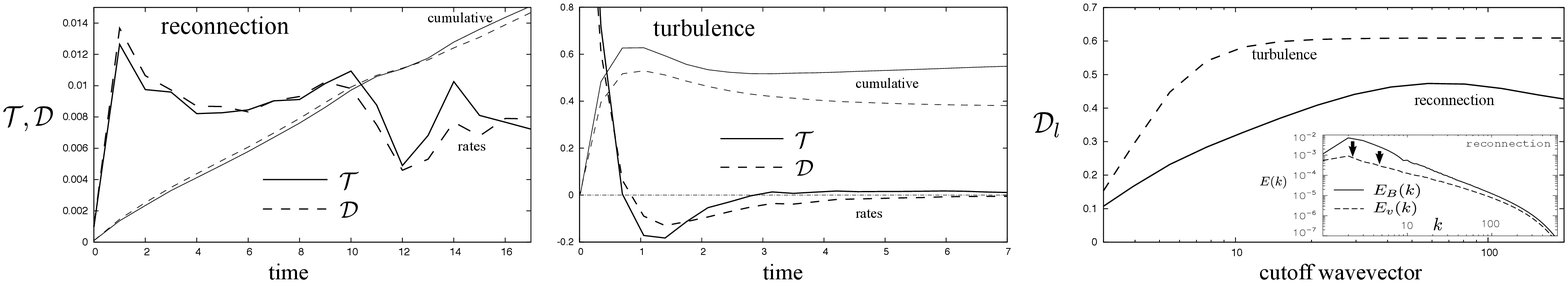}
\caption{A case study of terms related to curvature drift acceleration and energy conversion in spontaneous reconnection and decaying MHD turbulence.   
Left panel: case A, volumetrically averaged energy conversion rate $\cal T$ and curvature acceleration rate $\cal D$ in MHD simulation with turbulent current layer produced by spontaneous reconnection with setup described in \citet{B13a}. Right panel: case B, the same for decaying turbulence generated by random initial field. 
On both panels we also show cumulatives $\int {\cal \{T,D\}} dt$. The reconnection case (left panel) 
is characterized by approximately constant turbulent dissipation rate and it also show stable rate of energy conversion $\cal T$, while $\cal D$ closely follow $\cal T$. The decaying turbulence (middle panel)
shows a burst of energy conversion rate within a few dynamical times (Alfv\'enic times).
In both cases the gradient drift acceleration term (not shown) is relatively negligible.
Bottom panel shows the scale-dependency of the $\cal D$ term, by plotting ${\cal D}_l$ which is obtained by
calculating $\cal D$ with a coarse-grained dynamical quantities ${\bf v,B}$ (Gaussian low pass filter in Fourier
space with a cutoff wavenumber $k=2\pi/l$). ${\cal D}_l$ has a physical meaning of all energy transfer $B$ to $v$ accumulated down to scale $l$, for which reason it asymptotes to a constant at small $l$ or large cutoff wavenumbers. The inset in the right panel shows kinetic and magnetic spectra in case A to demonstrate
the range of scales within which  magnetic energy is transferred to kinetic -- down to $k\approx 30$ with thick arrows depicting the energy transfer.}
\label{terms}
\end{figure*}


\section{Curvature drift} 
To explore the implications of magnetic energy transfer in non-thermal particle acceleration,
it is instructive to consider the particles motion in slowly-varying
electric and magnetic field, which can be described in the so-called drift approximation.
The leading drift terms are known as electric, gradient and
curvature drifts. While electric drift, proportional to ${\bf [E\times B]}$, can not produce acceleration, the other two drifts can.
For example, imagine the configuration of the reconnection with the inflow, Fig~1 top panel. The gradient drift $\sim {\bf [B\times \nabla} B]$ is along $-z$,
and so is the electric field in the ideal case $E=-{\bf [u\times B]}/c$. Their product will be positive and will result in acceleration which is due to particles being compressed by the converging inflow. This mechanism does not work in the
initial, most energetic stages of spontaneous reconnection which has negligible inflow,
Fig.~1 middle panel
\citep{B13a}. It is this initial stage that has the highest volumetric dissipation rate, however. Fig~1 illustrates why curvature drift acceleration is important in this configuration. It also turns out that in any magnetically-driven turbulent environment, such as depicted on the bottom panel of Fig~1, curvature drift acceleration will accelerate particles on average.

Let us look carefully at the term which is responsible for acceleration by curvature drift,
\begin{equation}
\frac{d{\cal E}}{dt}=-2\frac{\cal E_\|}{B} \bf [u\times B] \cdot [b \times (b\cdot \nabla)b],
\label{curv_acc}
\end{equation}
 see, e.g., \cite{Sivukhin1965},
where ${\cal E_\|}=v_\| p_\|/2=\gamma m v_\|^2/2$ is a particle's parallel kinetic energy.
With some manipulation, this expression could be equivalently transformed as 
$2({\cal E_\|}/B) \bf u \cdot (B \cdot \nabla)b$. It now becomes clear that this term is related to the transfer rate between
magnetic and kinetic energies, in particular it is a fraction of $\cal D$:

\begin{equation}
 \frac{d{\cal E}}{dt}={\cal E_\|} \frac{8\pi}{B^2} \cal D.
\end{equation}

The physical meaning of this equation is that, given efficient particle scattering, so that ${\cal E_\|}={\cal E}/2$, the acceleration rate is determined by the
half of the local energy transfer rate $8\pi {\cal D}/B^2$, not including the $\cal X$ term.

\section{A case study} 
We can test the general ideas outlined above in two physical cases that feature turbulent energy transfer from magnetic to kinetic energies.  
Using spontaneous reconnection and the decaying MHD turbulence simulated numerically, we can directly calculate the discussed terms and compare them.
The spontaneous reconnection numerical experiment was started with thin planar current sheet and small perturbations in ${\bf u}$ and ${\bf B}$ and was described in detail in \citet{B13a}, while the decaying MHD turbulence was similar to our previous incompressible
driven simulations in \citet{BL09b}, except that there was no driving and the initial conditions were set as a random magnetic field with wavenumbers $1<k<5$ and zero velocity.
Both simulations developed magnetically-driven flows, from which we calculated the average $\cal T$, $\cal D$, and $\cal X$ terms and presented them in Fig.~2.

The spontaneous reconnection case had fairly stable reconnection rate, this also corresponded to the approximately constant $\cal D$ integrated over the volume.
The $\cal X$ term didn't seem to be sign-definite and contributed relatively little.
Gradient drift acceleration was also negligible, possibly due to the absence of global
compression.
Keeping in mind that all the energy had to come from magnetic
energy, and given that the dissipation rate was approximately constant, 
it was no surprise that the average $\cal D$ term evolved relatively little.
It should be noted, however, that in the spontaneous reconnection experiment the width of the reconnection region was growing approximately linear with time \citep{B13a},
so the $\cal D$ term magnitude, pertaining to the reconnection region itself was much higher than that of an averaged $\cal D$ over the 
total volume.
Given the reconnection layer thickness $l(t)$, the local
$\cal D$ could be estimated as $(1/2) v_r (B^2/8\pi)/l(t)$, where $v_r$ is a reconnection rate, which was $v_r \approx 0.015 v_A$ in \citet{B13a} and the $1/2$ comes
from only half of magnetic energy being transferred into kinetic energy before physically dissipating on very small dissipation scale $\eta \ll l(t)$. This would correspond to
acceleration rate of $(1/4) v_r /l(t)$ and can be very high, because the $l(t)$ could be as small as the Sweet-Parker current sheet width or the ion skin depth.

The decaying MHD turbulence experienced two regimes: 1) the initial oscillation when excessive amount of magnetic energy
was converted into kinetic energy and the bounce back and partial inverse conversion afterwards; 2) the self-similar decay stage of MHD turbulence.
The first stage had the strongest conversion term which was dominated by $\cal D$. All terms integrated over time were mostly accumulated
within the first 1-2 Alfv\'enic times.

\section{Scale-filtered quantities}  
It is known from turbulence theory that the energy transfer rate $\cal T$ can be demonstrated to be local in scale, under relatively weak assumptions \citep{Aluie2010,B12a}. The scale-locality means that each scale contributes
to the transfer independently. We also know empirically that most of the transfer between magnetic and kinetic energies happens on relatively large scales which are
comparable to the outer scale of the system, while below outer scale there is little
average transfer due to an approximate equipartition between kinetic and magnetic
energies.
For example, the reconnecting turbulent current layer has most of its $\cal T$ transfer within a factor of a few of the scale of the layer thickness,
while in the decaying turbulence problem it is within a factor of a few from the outer scale of the system.
Let us designate ${\cal T}_l$ as a transfer calculated from quantities which were filtered by low-pass Gaussian filter with a cutoff wavenumber of $1/l$.
Keeping in mind of locality we will conclude that only scales larger than $l$ will contribute to ${\cal T}_l$. This means that ${\cal T}_l$ will be constant
for small $l$ and will start decreasing when $l$ approaches the outer scale of the problem $L$. 
In general, we can not deduce the same for ${\cal D}_l$ and ${\cal X}_l$, but keeping in mind that ${\cal X}_l$ contributes relatively little
in two cases that we considered, ${\cal D}_l \approx {\cal T}_l$ in those cases. Fig.~2 demonstrates this behavior on the bottom panel, where energy transfer
is operating between wavenumbers 2 and 20 in the reconnection case (1/20 is approximately the layer width at this point), and the ${\cal D}_l$ mostly changes within the same range of scales.
The decaying turbulence case has the transfer more localized around the outer scales. 

In terms of drift, the particle with Larmor radius $r_L$ will ``feel'' magnetic and electric fields on scales larger
than $r_L$, while the scales smaller or equal to $r_L$ will contribute to particle scattering. It follows
that the ``effective'' ${\cal D}$ will be ${\cal D}_{r_L}$, an implicit function of energy.
Combining this with the result obtained above that ${\cal D}_l$ goes to a constant for small $l$ we conclude
that the acceleration rate will also go to a constant for particles with $r_L$ smaller than the system size.
A similar, more hand-waving argument, is that the term ${\cal D}_l$ could be roughly
approximated as $B_l^2 v_l/l$, resembles turbulent energy transfer rate, which is scale-independent. Interestingly, starting with
scale independent energy transfer in turbulence we arrived at the energy-independent acceleration rates. Given the generality of the arguments
presented above it is not surprising that energy-independent rates were indeed observed in simulations \cite{Guo2014}.

\section{Acceleration in spontaneous reconnection} 
The development of the thin current sheet instability results in turbulence and reconnection in a sense of dissipated magnetic energy. This process will come through two distinct
regimes, the regime without significant outflow for times smaller than $L/v_A$ \citep{B13a} and the stationary reconnection with outflow for larger times \citep{Lazarian1999}.
Let us consider the first regime which has larger dissipation rate per unit volume, because the current layer thickness $l(t)=v_r t$ is relatively small. We use $v_r$ for
the reconnection rate and $t$ as the time since the beginning of spontaneous reconnection. Let us assume that the current layer thickness is limited from below by the
ion skin depth. We will have acceleration rate of $1/(4t)$ for all times larger than $d_i/v_r$ but smaller than $L/v_A$. The solution for energy, therefore, will
be ${\cal E}={\cal E}_0 (t v_r/d_i)^{1/4}$ where ${\cal E}_0$ is the initial energy, e.g., the thermal energy. The particle's energy will be
${\cal E}_{max}=0.35 {\cal E}_0 (L/d_i)^{1/4}$ given the reconnection rate $v_r=0.015 v_A$. This, however, is only the maximum energy of accelerated particles, as only
a tiny fraction of particles were contained in the original thin current sheet and started accelerated from initial time $d_i/v_r$ \citep{Guo2014}. With proper stochastic treatment
and assuming that the escape rate $r_{esc}$ is negligible compared with acceleration rate in this no-outflow problem, we expect to see the power law particle spectrum
with the $-1$ slope and a cutoff at ${\cal E}_{max}$. 

The subsequent development of an outflow will do three things: first it will enable the inflow and therefore the extra acceleration term associated with gradient drift or converging
magnetic mirrors \citep{dalpino2005}. Secondly, it will stabilize acceleration rate for the curvature drift acceleration at $v_A/4L$, as the current layer is no longer expanding. Thirdly, it will enable
particle escape through the outflow. In this regime the spectrum will extend from ${\cal E}_{max}$ to higher energies, up to Larmor radii of the large scale
of the current sheet $L$. The spectral slope of this extension will be determined by $-1-r_{esc}/r_{acc}$, where $r_{acc}$ should account for all acceleration and cooling
mechanisms, such as gradient drift acceleration and outflow cooling. The detailed analysis of this stage will be the
subject of a future work. 

The electron spectra observed in solar X-ray flares are fitted with the thermal component with temperature of several keV and the steep power law component.
This is consistent with our picture, as the rather shallow $-1$ slope, containing most of its energy near ${\cal E}_{max}$, is likely to thermalize. Also, the outflow phase will extend this distribution as a power-law to higher energies. 

\section{Discussion}  
We demonstrated that magnetic configurations that relax to the lower states of magnetic energy will also regularly accelerate
particles, on timescales which are, typically, Alfv\'enic, but can be much shorter, e.g., in the beginning of spontaneous reconnection.
This mechanism of acceleration of collisionless nonthermal particles by MHD electric field
should not be confused with the acceleration of the bulk of the plasma by magnetic tension. Indeed,
for partiles with low energies the drift terms could be neglected, i.e. expression (\ref{curv_acc}) will
trivially turn to zero. In the bulk fluid acceleration the energy gained by each particle does not depend
on its initial energy, while for dift acceleration scenario it is proportional to the particle energy.

Some recent observations, e.g., \citep{aharonian2007,abdo2011,aleksic2011} suggested 
high energy emission variability could be as fast as variability at lower energies,
which is at odds with DSA, which predicts acceleration timescale proportional to diffusion
coefficient which is typically a positive power of energy. This has been pointed out as a moivation for
reconnection scenario \citep{Clausen2012,Giannios2013,Zhang2011}.

Particle acceleration during reconnection is a topic under intense study, but the mechanism discussed in this paper is distinctly different
from the direct acceleration by the reconnecting electric field at the X-line. In fact, we completely ignore non-ideal effects which produce local $\bf E \| B$. Also, our mechanism is not tied to a special X-point, but instead volumetric. An interesting first-order acceleration mechanism in ideal MHD turbulence related to imbalanced turbulence and convergent field lines has been suggested recently by \citet{Schlickeiser2009}.

The regular acceleration due to converging field lines have been suggested in \citet{dalpino2005}, later it was pointed by \citet{drury2012a} that an
outflow cooling should also be included. In this paper we do not rely on simple transport equation, such as Parker's, therefore we relax the above approach's requirement that particles need to be almost isotropic.
The acceleration in turbulent reconnection has been further numerically studied in \citet{kowal2012b}. 
Plasma PIC simulations has also been increasingly used to understand particle acceleration. The emphasis was mostly on the non-ideal effects
near X-line regions and interaction with magnetic islands \citep{zenitani2001,pritchett2006,fu2006,drake2006,oka2010,dahlin2014}. The change of energy due to curvature drift in a single collision of a particle with magnetic island was estimated analytically in \citet{drake2006}. An important question that was left out in that study was whether
this term will result in acceleration or cooling, on average. Without understanding this, it was not clear whether this process results in acceleration or deceleration or the second-order
effect. In this paper we unambiguously decide this by relating the answer to a certain well-known
statistical measure in turbulence -- the direction of transfer between magnetic and kinetic energies. 
We also showed that the curvature drift acceleration does not require particles to be trapped in contracting magnetic islands, so their energy is not limited by this requirement, meaning that the
energy cutoff is not related to the island size, instead it is related to the system size, see also \citet{Uzdensky2014}.

PIC simulations are limited in the range of scales and energies they cover. Recent simulations in \citet{Sironi2014,Guo2014} demonstrated acceleration up to 100 MeV in electron energies, which is below maximum energy in most astrophysical sources. Theory, therefore, is necessary to supplement conjectures based on the observed PIC distribution tails, explaining
the underlying physical mechanism and making predictions for astrophysical systems which often feature gigantic scale separation between
plasma scales and the size of the system. The feedback from simulations, nevertheless, was very useful, in particular the recent simulations \citep{Guo2014} that reached MHD scales and confirmed the prediction that the curvature drift acceleration will dominate compared to the non-ideal electric field acceleration. 

\section{Acknowledgements} 
We are grateful to Fan Guo for sharing his results on a preliminary stage. The work is supported
by the LANL/LDRD program and the DoE/Office of Fusion Energy Sciences through CMSO.
Computing resources at LANL are provided through the Institutional Computing Program.
We also acknowledge support from XSEDE computational grant TG-AST110057.


\def\apjl{{\rm ApJ }}          
\def\grl{{\rm GRL }}
\def\aap{{\rm A\&A } }
\def\mnras{{\rm MNRAS } }
\def\physrep{{\rm Phys. Rep. } }               

\bibliography{all}

\end{document}